\documentclass[aps,pre,twocolumn,preprintnumbers,amsmath,amssymb]{revtex4}
\input epsf \usepackage[pdftex]{graphicx}
\usepackage{dcolumn}
\usepackage{bm}

\begin{document}

\title{Three-dimensional "Mercedes-Benz" model for water} 

 \author{Cristiano L. Dias$^{1}$} \email{diasc@physics.mcgill.ca}
\author{Tapio Ala-Nissila$^{2,3}$}\email{Tapio.Ala-Nissila@tkk.fi}
\author{Martin Grant$^{4}$} \email{martin.grant@mcgill.ca}
\author{Mikko Karttunen$^{1}$} \email{mkarttu@uwo.ca}
\affiliation{$^{1}$Department of Applied Mathematics,
The University of Western Ontario, London, Ontario, Canada N6A\,5B7\\  
$^{2}$Department of Physics, Brown University, Providence RI 02912-1\,843 \\
$^{3}$COMP Center of Excellence and Department of Applied Physics, Helsinki University of Technology,
P.O. Box 1100, FI-02015 TKK, Espoo, Finland \\
$^4$Department of Physics, McGill University,
3600 rue University, Montr\'eal, Qu\'ebec, Canada H3A\,2T8}

\date{\today}

\begin{abstract}
In this paper we introduce a three-dimensional version of the
Mercedes-Benz model to  describe water molecules. In this model van
der Waals interactions and hydrogen bonds are  given explicitly
through a Lennard-Jones potential and a Gaussian orientation-dependent
terms, respectively.  At low temperature the model freezes forming Ice-I
and it  reproduces the main peaks of the experimental radial
distribution function of water. In addition to these structural
properties, the model also captures the thermodynamical
anomalies of water: the anomalous density profile, the negative
thermal expansivity, the large heat capacity  and the minimum in the
isothermal compressibility.
\end{abstract}


\maketitle

\section{Introduction}

Water is the most important fluid on earth. It covers two thirds of
the planet's surface and controls its climate.  Most importantly,
water is necessary for carbon-based organic life being the solvent 
in most \textit{in vivo} chemical reactions. 
Its unique hydration  properties drive biological macromolecules towards
their three-dimensional  structure, thus accounting for their function
in living organisms~\cite{Chaplin:06aa}. Water exhibits anomalous properties that 
affect life at a larger scale. For example,
mammals benefit from the large latent heat of water to cool them down
through sweating,  while water's large heat capacity prevents local
temperature fluctuations, facilitating  thermal regulation of
organisms. 

These anomalous properties result from a competition between
isotropic van der Waals interactions and  highly directional 
hydrogen bonding (H-bond). 
A large number of models of varying complexity 
have been developed and analyzed 
to model water's extraordinary properties, for reviews see e.g. Refs.~\cite{Jorgensen:83aa,Nezbeda:97aa,Guillot:02aa,VEGA09},
but none of the current models can correctly reproduce all physical properties of water. 
Those model are typically calibrated against experimental data, for 
example the radial distribution function (RDF) at ambient 
conditions~\cite{SORE00,SOPE00}, or the temperature of maximum density~\cite{MAHO00}, 
i.e. $T=3.98^\circ$C. 
While there is no guarantee that a 
model optimized to reproduce a given property is able to account 
for others, adding details increases its quantitative accuracy. 
For example, TIP5P, which describes water through 5 interacting sites, is typically
more accurate~\cite{MAHO00} than models with 3 or 4 interacting sites.  
The addition of each interacting site, however, makes the model 
considerably much more demanding computationally. This is an undesirable feature 
since a large number of 
water molecules is required to hydrate even the smallest peptides resulting in a high 
computational cost. Thus, simple models, such as SPC~\cite{Berendsen:81px} 
and TIP3P \cite{Jorgensen:83zc,Neria:96sc}, are the most  
used ones in computational studies of  biologically motivated systems. 
In addition, new models and improvements appear frequently in literature, 
see e.g., Refs.~\cite{price:04aa,horn:04aa,Wu:06xx,baranyai:07aa}
and references in them.

Coarse-grained models have also been developed and  used to study the emergence of water's 
anomalous properties from its atomic constituents. 
Both lattice~\cite{BARB08,BUZA04,PRET05} and continuous 
models~\cite{SILV98,POOL94,Lyubartsev:00uc,Lyubartsev:02nz} have been applied. 
Current coarse-grained models cannot, however,
be easily used to study hydration 
of macromolecules, because they can not reproduce the structure of liquid water
which is essential in studies of biological systems and molecules~\cite{Chaplin:06aa}. 
A proper structural description is required since hydration and, in particular, the 
hydrophobic effect, which is the main driving force for protein folding~\cite{KAUZ59,DILL90}, 
depend on the amount of structural order close 
to the hydrated molecule versus the amount of order in bulk water. A simplified model 
that would account for both thermodynamical and structural properties of water, 
would therefore be highly beneficial in studies related to the hydrophobic 
effect, protein folding and macromolecules in general. 

The main purpose of this work is to introduce a simple but realistic
model that reproduces both the main structural and thermodynamic
properties of water. To this end, we extend the two dimensional (2D)
Mercedes-Benz (MB) model~\cite{NAIM71}  to 3D. In 2D, the MB model
has already provided insights into several properties of water:~its
anomalous thermodynamical behavior \cite{SILV98},  hydration of
non-polar solutes~\cite{SOUT02},  ion solvation~\cite{DILL05}, cold
denaturation of proteins~\cite{DIAS08},  and the properties of
different amino acids \cite{BECK06}. Despite this success,  there are several
mechanisms which cannot be studied in two dimensions and an extension
to 3D is needed.
%

We show that a previously proposed framework for the 3D MB model~\cite{BIZJ07}
does not reproduce the thermodynamical anomalies of water.  Here, we
extend the model to overcome that problem by making H-bonding
dependent on the local environment of atoms by
penalizing compact  configurations in favor of open-packed ones. With
this implementation,  structural and thermodynamical properties of
are recovered  qualitatively. We would like to emphasize that
our goal is not to replace atomistically accurate models, 
such as TIP5P, but rather to provide an alternative for qualitative studies
involving water.

The rest of this paper is organized as follows: next, we introduce 
and discuss the existing 3D MB model, and propose a correction that makes
the model suitable to describe thermodynamical and structural 
properties of water. In the same section, we present 
the Monte-Carlo scheme and the cooling protocol used in this work. 
We present our results and comparison to experimental results in
Sec.~\ref{sec:results}.
In the section entitled results, experimental data are compared 
qualitatively to our simulations. 
Finally, we present our conclusions and a discussion in Sec.~\ref{sec:concl}.

\section{The model}

\subsection{Mercedes Benz model}

\begin{figure}
\centerline{\includegraphics[height=40mm]{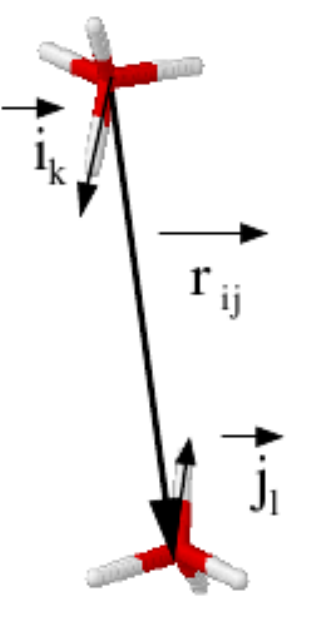}}
\caption{(Color online) Schematic representation of two Mercedes-Benz molecules and important vectors defining their interaction. }
\label{fig:MB}
\end{figure}

In the 3D MB model, water molecules interact explicitly through two types of 
empirical potentials: H-bonds and van der Waals. H-bonds are directional and account 
for the tetrahedral structure of water which is described by four arms 
separated from each other by angles of $109.47^\circ$, see Fig.~\ref{fig:MB}. 
The energy of H-bonds is minimized
whenever arms of adjacent molecules point towards each other. Mathematically if $\vec{X_i}$ 
represents the position of the $i^\mathrm{th}$ particle and its four unitary arms, which 
are denoted by $\vec{i_k}$ 
(with $k=1,2,3,4$), then H-bond interaction between molecules $i$ and $j$ can be written as
\begin{equation}
U_{HB}(\vec{X_i},\vec{X_j}) = \sum_{k,l=1}^4 U_{HB}^{kl}(r_{ij},\vec{i_k},\vec{j_l}),
\end{equation}
where
\begin{eqnarray}
U_{HB}^{kl} (r_{ij},\vec{i_k},\vec{i_l})  & = & 
\epsilon_{HB}G(r_{ij}-R_{HB},\sigma_R)  \times \nonumber \\
& \, &  G(\vec{i_k}\hat{r}_{ij}-1,\sigma_{\theta})
G(\vec{j_l}\hat{r}_{ij}-1,\sigma_{\theta})
\label{eqn:Hbond}
\end{eqnarray}
and $G(x,\sigma)$ is an unnormalized Gaussian function
\begin{equation}
G(x,\sigma) =  \exp [-x^2/2\sigma^2].
\end{equation}
The above mathematical description ensures that the intensity of a H-bond is maximized 
whenever the arms of neighboring molecules are aligned with the vector $\vec{r_{ij}}$ joining
their centers of mass and whenever their distance is equal to $R_{HB}$.

The spherically symmetric van der Waals interactions are approximated by a Lennard-Jones potential:
\begin{equation}
U_{LJ}(r_{ij})= 4 \epsilon_{LJ}\Big[ \Big(\frac{\sigma_{LJ}}{r_{ij}}\Big)^{12} - 
\Big( \frac{\sigma_{LJ}}{r_{ij}}\Big)^{6}\Big],
\end{equation}
where $\epsilon_{LJ}$ describes the strength of the interaction and $\sigma_{LJ}$ is the 
particle diameter.
Then, the total energy describing two MB particles is given by
\begin{equation}
U(\vec{X_i},\vec{X_j}) = U_{LJ}(r_{ij})+U_{HB}(\vec{X_i},\vec{X_j}).
\end{equation}

Bizjak \textit{et al.}~\cite{BIZJ07} studied this model using the
following set of   parameters:~$\epsilon_{HB}=-1$,
$\epsilon_{LJ}=1/35\epsilon_{HB}$, $R_{HB}=1$,  $\sigma_{LJ}=0.7$,
$\sigma_R = \sigma_{\theta} = 0.085$. They assumed a diamond structure
for the model's ground state which is the configuration taken by
oxygen atoms when water forms cubic-ice, i.e. Ice-I$_c$. When tested  
against simulation, however, this
assumption fails and the model can be shown to  minimize its energy in
an Ice-VII configuration:~two interpenetrating diamond lattices with
no H-bonds connecting these lattices. 

Ice-VII appears to be the optimized ground state for systems trying to
maximize their density within a tetrahedral symmetry. It is therefore
natural that the 3D MB model of Bizjak \textit{et al.}~\cite{BIZJ07}, 
whose H-bond term imposes a tetrahedral configuration and the van 
der Waals term favors compact 
conformations, has this structure as its  ground state. 
However Ice-VII is not the desired ground state for models of water 
at ambient pressure such that 3D lattice models for this material have an 
explicit energetic term penalizing compact 
configurations of this type~\cite{PRET05,ROBE96,BELL72}.

\begin{figure}
\vspace{-3cm}
\begin{tabular}{c}
\centerline{\includegraphics[height=145mm]{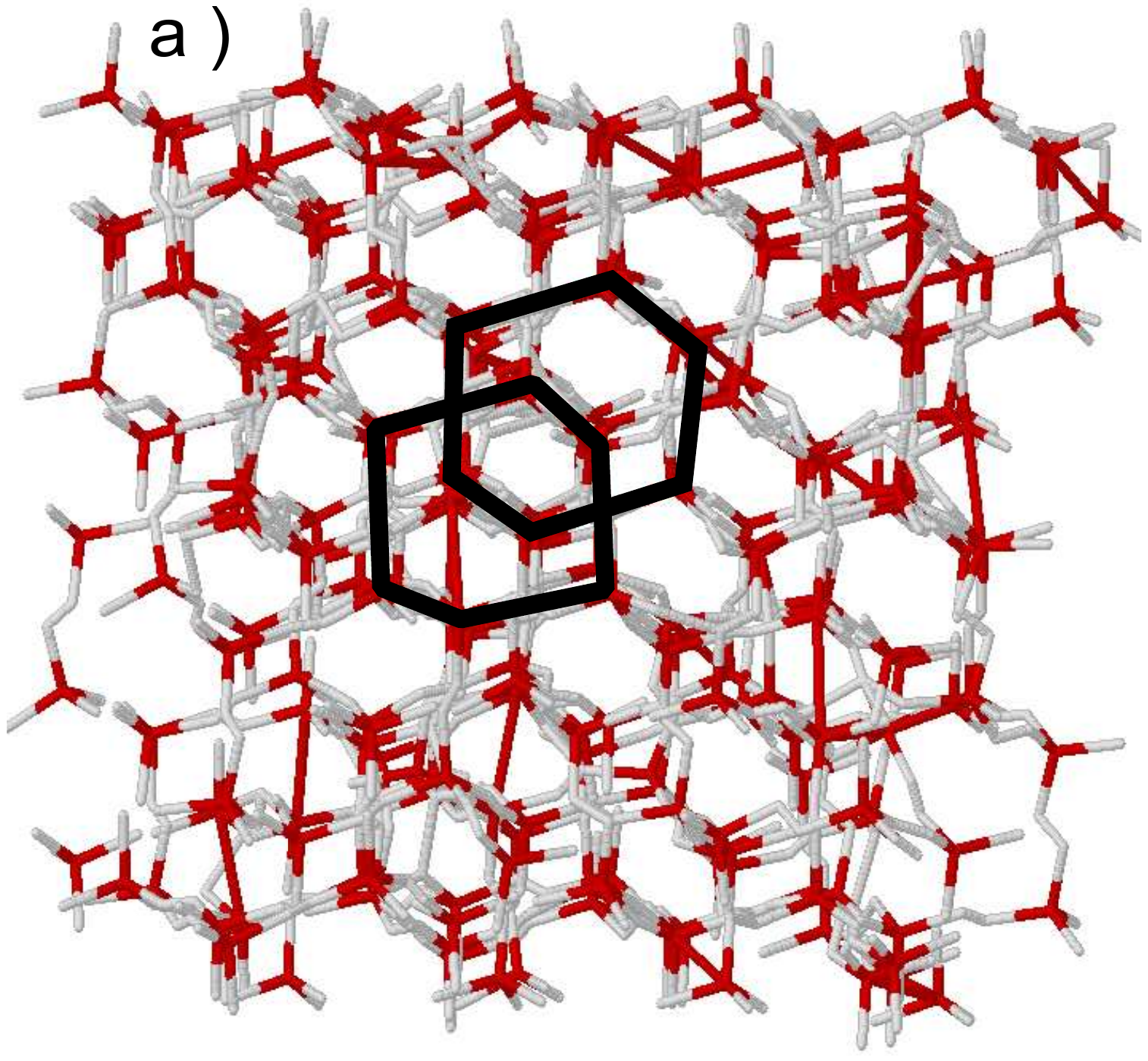} }  \vspace{-1.5in}\\
\centerline{\includegraphics[height=60mm]{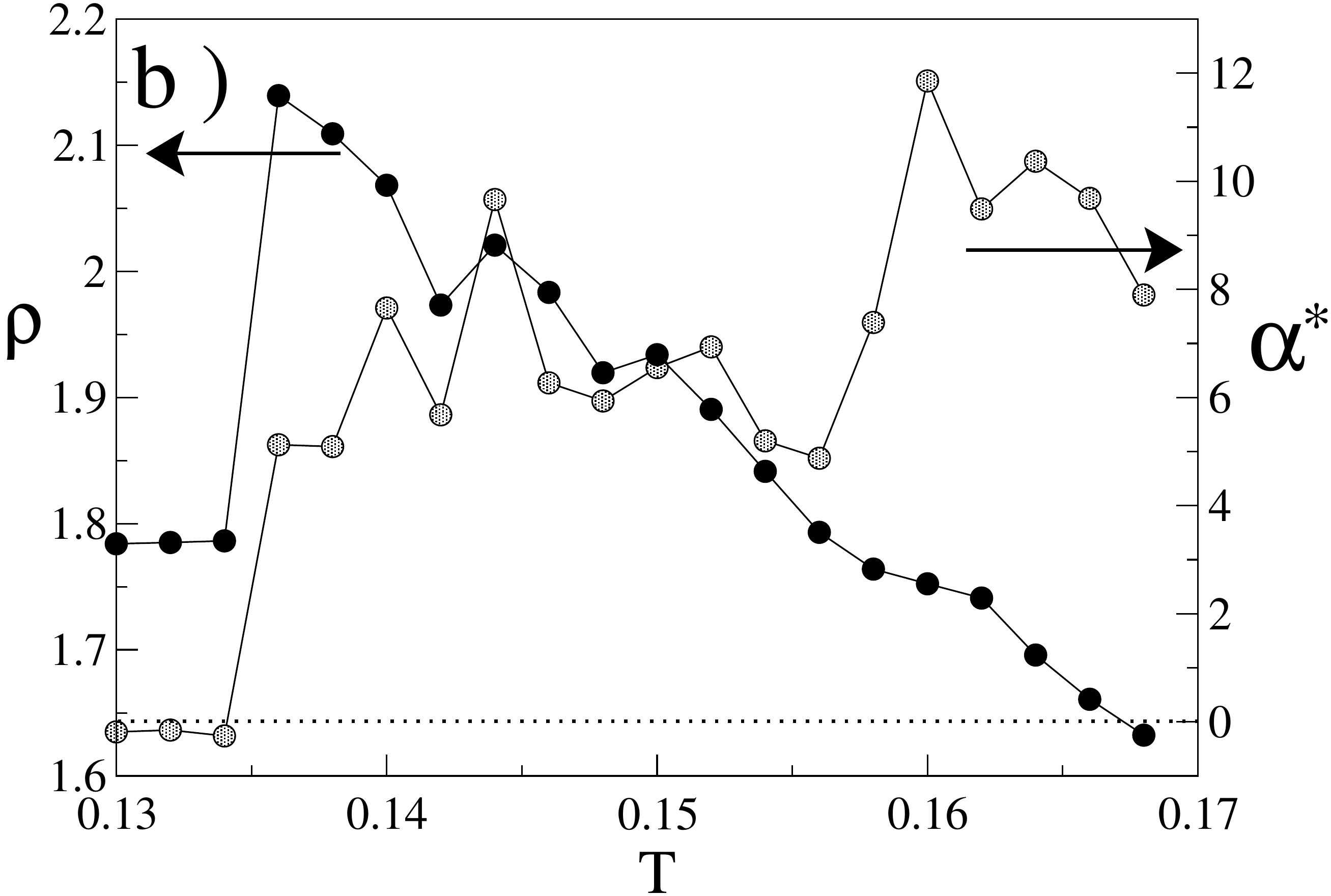} }
\end{tabular}
\caption{(Color online) (a) Ground state of the Mercedes-Benz model: Ice-VII. As a guide to the eye
two hexagons, representing the two interpenetrating diamond structure of ice-VII,
are drawn. (b -- left axis) Dependence of 
the density (g/cm$^3$) on temperature (in units of $\epsilon_{HB}$). 
(b -- right axis) Dependence of the coefficient of thermal expansion 
($\epsilon_{HB}^{-1}$) on temperature. A constant pressure 
of 0.2 in units of $\epsilon_{HB}/R^3_{HB}$ was used.}
\label{fig:dill}
\end{figure}

Figure~\ref{fig:dill}(a) shows a typical Ice-VII ground state obtained by quenching a system of 
256 MB particles interacting through the framework of 
Bizjak \textit{et al.}. Details about the simulation method and
the cooling procedure will be described later in this section. 
Figure~\ref{fig:dill}(b) shows the density obtained along quenching. The liquid phase has 
a higher density than ice -- as in real water. However, the model does not reproduce the 
density anomaly of water:~the liquid phase does not show a temperature of 
maximum density below which the density decreases. As a result, the thermal expansion 
coefficient is never found to be negative in the liquid phase (Fig.~\ref{fig:dill}(b)). 
In addition, the model does not reproduce the structure of liquid water (see Fig.\,3 in Ref.~\cite{BIZJ07}): 
the simulated RDF has a non-realistic peak at a distance corresponding
to the van der Waals radius.

Despite the problems cited above, the 3D MB model remains an attractive 
coarse-grained model for water. It does not require calculation of charges, which
enables longer simulation times  desperately needed in  studies of macromolecules. 
It also holds the promise of being able to provide a qualitatively accurate description 
of the structure of water due to its 
tetrahedral nature~\cite{BERN33} and water's thermodynamical properties since 
the model exhibits both open and close packed structures required to described water's
anomalous behavior~\cite{POOL94}. Next, we describe how the model of 
Bizjak \textit{et al.}~\cite{BIZJ07} can be 
improved to better describe the structure and thermodynamics of water.

\subsection{Corrections to the Mercedes-Benz model}

To resolve the above problems, we introduce a term that depends
on the local environment of particles. Our approach is inspired by Tersoff-like potentials for 
covalent materials~\cite{TERS88}. This term penalizes H-bonds which are formed in crowded
environments through the factor
\begin{equation}
b(z_{i})=
\begin{cases} 
1 ,& if z_{i} \leq 4 \\
\Big( \frac{4}{z_{i}} \Big)^\upsilon, & if z_{i} > 4, 
\label{eqn:penalty}
\end{cases}
\end{equation} 
where $z_{i}$ is the coordination of atom $i$, computed as 
$z_{i}~=~\sum_{k \ne i} f( r_{ik} )$ with the cut-off function defined by~\cite{TERS88}:
\begin{equation}
f(r_{ij})=
\begin{cases}
1, & r < R-D \\
\frac{1}{2}-\frac{1}{2}\sin\Big( \frac{\pi}{2}(r-R)/D \Big) ,& R-D<r<R+D \\
0, & r>R+D
\end{cases}
\end{equation}
where $R$ and $D$ are chosen as to include the first-neighbor shell only. 
Note that $f(r)$ decreases continuously from 1 to 0 in the range $R-D<r<R+D$.

The energy of H-bonds corrected through Eq.~(\ref{eqn:penalty}) becomes:
\begin{equation}
U^c_{HB}(\vec{X_i},\vec{X_j}) = b(z_i) \sum_{k,l=1}^4 U_{HB}^{kl}(r_{ij},\vec{i_k},\vec{j_l}).
\end{equation}
This equation penalizes H-bonds whenever interacting molecules have more than 
four neighbors. This inhibits the formation of compact tetrahedral 
phases, e.g. Ice-VII, and favors open-packed tetrahedral phases such as Ice-I. 

In order to ensure that H-bonds favor chair-like configurations required for diamond 
structure, we also add a standard potential with three-fold symmetry for dihedral angles. 
This potential adds an energetic cost to the H-bond between arms $m$ of molecule $i$ 
and arm $n$ of molecules $j$, if the dihedral angle formed by the other arms of these molecules is not $60^{\circ}$:
\begin{eqnarray}
U_{\phi}^{mn}(\vec{X_i},\vec{X_j})=
\frac{\epsilon_{\phi}}{2} U_{HB}^{mn}(r_{ij},\vec{i}_m,\vec{j}_n)b(z_i) \times \\ \nonumber
\sum_{k \ne m \atop l \ne n}( 1 + \cos(3\phi_{kl}) ),
\label{eqn:dihedral}
\end{eqnarray}
where $\epsilon_\phi$ is the strength of the interaction. The term 
$U_{HB}^{mn}(r_{ij},\vec{i}_m,\vec{j}_n)b(z_i)$ 
ensures that the penalty is proportional to the strength of the H-bond.
The dihedral angle $\phi_{kl}$ describes how the arm $k$ of molecule $i$ aligns 
with the arm $l$ of molecule $j$ along the vector joining the center of mass of 
these two molecules. Thus, the total dihedral energy between molecules $i$ and $j$ is
\begin{equation}
U_{\phi} (\vec{X_i},\vec{X_j})= \sum_{m,n} U_{\phi}^{mn}(\vec{X_i},\vec{X_j}).
\end{equation}

Note that because of the dependence on the local environment, 
$U^c_{HB}(\vec{X_i},\vec{X_j}) \ne U^c_{HB}(\vec{X_j},\vec{X_i})$ and 
$U_{\phi}^{mn}(\vec{X_i},\vec{X_j}) \ne U_{\phi}^{nm}(\vec{X_j},\vec{X_i})$. This 
asymmetry has no physical implications since $U^c_{HB}$ and $U_{\phi}$ 
possess all the invariance properties required for a potential~\cite{TERS88}. 

We can now write the total potential energy between two water molecules as
\begin{equation}
E(\vec{X_i},\vec{X_j})\,=\,U_{LJ}(r_{ij})\,+\,U^c_{HB}(\vec{X_i},\vec{X_j})\,+\,U_{\phi}(\vec{X_i},\vec{X_j}).
\end{equation}
This model has 10 parameters which were chosen such as to account for a
semi-quantitative agreement of the density profile with experiment.
We proceeded in two steps to adjust these parameters. First, we chose 
the values for these parameters such as to produce a density 
in g/cm$^3$ \footnote{The density can  be
computed in g per cm$^3$ by mapping $R_{HB}$ to  its
experimental value~\cite{SOPE00}, i.e. ~$R_{HB}=2.78$ \AA, and using
$M=2.992\cdot 10^{-23}$ g for the molecular mass of water:~$
\rho  = \Big( \frac{256}{V} \Big) 1.45448~\mathrm{g/cm}^3$.} that is
comparable to  experimental values, i.e. about 1 g/cm$^3$  for the
liquid phase and 0.93 g/cm$^3$ for the ice phase. Only under this
condition can the structure of the model be qualitatively
similar  to real water. Then, we adjusted the parameters such as to
obtain a density that is a concave function of temperature with its
maximum close to the freezing point. This second condition is the
minimal requirement for describing the anomalous properties of water.

The set of parameters calibrated according to the above 
procedure is given here in reduced units. We 
report energies and distances in terms of the binding energy  $|
\epsilon_{HB}|$ and equilibrium distance $R_{HB}$ of the H-bond.  In
these units, the three binding energies describing the system are 
$\epsilon_{HB}=-1$, $\epsilon_{LJ}=0.05$ and $\epsilon_{\phi} =
0.01$. The two distances are $R_{HB}=1$ and $\sigma_{LJ}=1.04 /
2^{1/6}$. The two terms controlling H-bond
interaction are $\sigma_R = 0.1$ and $\sigma_{\theta}=0.08$, and the
three parameters controlling the penalty of crowded environments are
$\upsilon = 0.5, R=1.3$ and $D=0.2$.  In this work, temperature is
given in units of $|\epsilon_{HB}|/k_B$, where  Boltzmann's
constant $k_B$ is set to unity. Pressure is given in units of
$|\epsilon_{HB}|/R^3_{HB}$. 

While adjusting the parameters, we found that the behavior of the
system is robust upon changing the  variables characterizing crowded
environments. It is, however, sensitive  to the ratio between the
binding energy of the van der Waals interaction and the binding energy
of the H-bond. This ratio controls the interplay of forces leading to
an environment where MB molecules are radially  surrounded by their
first-neighbors, and forces favoring a tetrahedral distribution of
the first-neighbors. The latter favors a high density configuration while
the former accounts for  a low density one. As opposed to the 2D MB model,
we kept the equilibrium distance of the van der Waals interaction comparable to
the equilibrium distance of the H-bond such as to avoid artificial peaks in the RDF~\cite{BIZJ07}.

\subsection{Numerical simulation method}
\begin{figure*}
\centerline{\includegraphics[height=120mm]{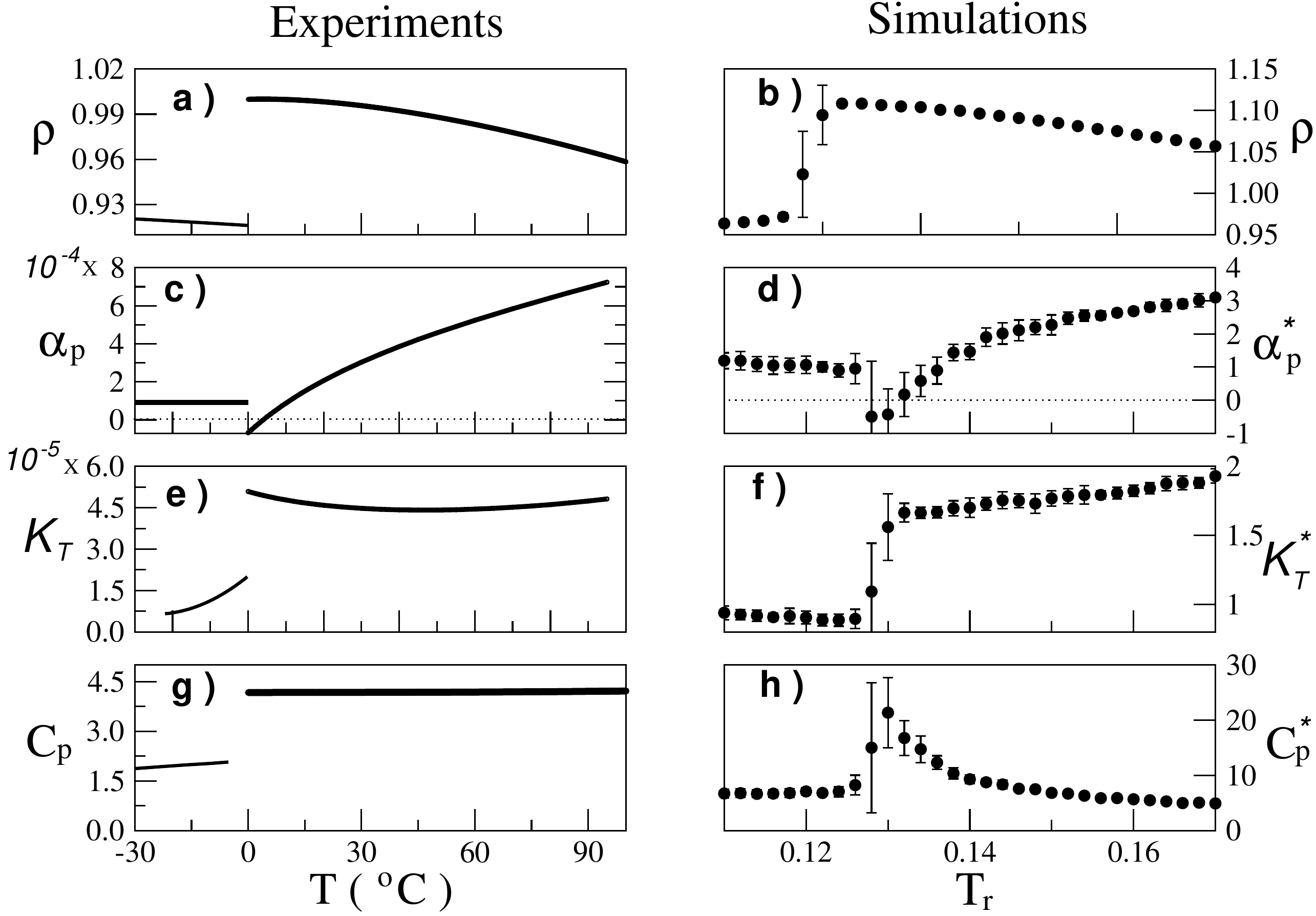} }
\caption{Thermodynamical properties of water. \emph{Left:} Experiments -- 
data obtained from references~\cite{KELL75}. Units are: $\rho$ (g/cm$^3$), $\alpha_P$ (K$^{-1}$), 
$C_P$ (J / g / K) and $\kappa_T$ (bar$^{-1}$). \emph{Right:} Simulations. Units are:~
$\alpha^*_P$ ($\epsilon_{HB}^{-1}$), $\kappa^*_T$ ($R_{HB}^3/\epsilon_{HB}$) and 
$C^*_P$ (dimensionless units). 
}
\label{fig:water}
\end{figure*}

For numerical simulations, we use the isothermal-isobaric (NPT) ensemble to study the thermodynamical 
properties of a system made of $N=256$ MB particles. A Monte-Carlo scheme is used where, 
at each step, an attempt is made to
displace the center of mass and the orientation of particles randomly by a 
quantity $\Delta R_{max}$ and $0.125$ rad, respectively. The maximum translational 
displacement is chosen such as to give an acceptance ratio of 50\%. Periodic 
boundary conditions are used to mimic an infinite system and at every 5 Monte 
Carlo sweeps, an attempt to rescale the size of the box is made (1 Monte Carlo sweep is 
equivalent to $N$ attempted steps).

To obtain thermodynamical data throughout the desired range of temperatures, the initial
configuration of the system is chosen randomly and equilibrated at the highest temperature 
($T=0.17$) for $5 \times10^4$ sweeps, after which statistics are gathered for the same
amount of time. Then, the system is cooled down by $\Delta T = 0.002$ and a 
similar cycle of equilibration/data gathering is performed. This cooling 
procedure is repeated until the lowest temperature, i.e. $T=0.11$ is reached. At the 
transition temperature an additional cycle of equilibration/statistics gathering ensured 
that the system was equilibrated properly. For all the pressures studied here, this protocol 
was repeated for 10 samples differing by the initial condition. All the 
quantities reported are the average over those 10 samples and, whenever relevant, the 
root-mean-square of this average is also shown as the error-bar.

The quantities computed during the simulations were the average potential energy per 
particle $E$, the volume per particle $V$, the heat capacity $C_P$, the 
compressibility $\kappa_T$ and the thermal expansion coefficient $\alpha_P$. The last three 
quantities are computed mathematically from the standard fluctuation relations:
\begin{eqnarray}
C^*_P &=& \frac{C_P}{k_B}=\frac{\langle H^2\rangle-\langle H\rangle ^2}{NT^2}, \nonumber\\ 
\nonumber\\
\kappa^*_T &=& \frac{\langle V^2\rangle-\langle V\rangle^2}{T\langle V \rangle},  \\ 
\nonumber \\
\alpha^*_P &=& \frac{\langle VH \rangle-\langle V \rangle \langle H \rangle}{T^2\langle V \rangle}, \nonumber
\end{eqnarray}
where $H$ corresponds to the enthalpy of the system. As for the other quantities computed
during the simulation, these response functions will be given in reduced units. Thus, $C^*_P$ will
be reported in dimensionless units, $\kappa^*_T$ in terms of $R_{HB}^3/\epsilon_{HB}$ and 
$\alpha^*_P$ in units of $\epsilon_{HB}^{-1}$.

\section{Results \label{sec:results}}

In Fig.~\ref{fig:water}, we provide a qualitative comparison between the 
properties of bulk water (left panels) and the MB model at $P=0.2$ (right panels). 
The behavior of the MB model follows the trends of water quite 
accurately: the anomalous density profile (panels on the first row), 
the negative thermal expansivity (second row), the minimum in the isothermal 
compressibility (third row) and the large heat capacity (fourth row).

At ambient pressure, water freezes into an open packed configuration called hexagonal-ice, i.e. 
Ice-I$_h$. This structure is held together by H-bonds which break when ice melts. 
At this transition, water molecules fill part of the empty spaces, assuming a higher 
density. In the liquid phase close to the melting temperature, 
a few open-packed configurations persist -- held together by H-bonds.
As the system is heated up, those bonds melt gradually removing empty spaces
and increasing the density of the system. This reduction of empty spaces occurs until 
the temperature of maximum density is reached. At this point thermal fluctuations decrease 
the density of the system 
with increasing temperature. This behavior has been measured experimentally
(Fig.~\ref{fig:water}(a)) and is captured by the MB model (Fig.~\ref{fig:water}(b)):~abrupt 
increase of the density at the melting transition and concave temperature dependence 
for the density of water with a maximum close to the melting transition. 

The thermal expansion coefficient is proportional to the derivative of the volume with 
respect to temperature $\alpha_P = 1/V (\partial V / \partial T)_P$. As for most materials, 
$\alpha_P$ decreases upon cooling (Fig.~\ref{fig:water}(c)) -- 
indicating that the volume of water decreases with temperature. 
It becomes zero at the temperature of maximum density and negative 
close to the freezing point. This unusual negative expansivity is reproduced in the model
(Fig.~\ref{fig:water}(d)) and reflects the unusual behavior
of water to expand upon cooling below the temperature of maximum density.

The isothermal compressibility measures the tendency of a system to
change its volume when the applied pressure is varied:~$\kappa_T = -
1/V(\partial V / \partial P)_T$. For a typical material, $\kappa_T$
decreases upon cooling since it is related to density  fluctuations
whose amplitude becomes smaller as temperature decreases. This is in
contrast with the behavior of water (Fig.~\ref{fig:water}(e)). The
compressibility of water  is a convex function of temperature and has
a minimum. This anomalous behavior can be explained  by noticing that the 
compressibility is lower for highly packed system than for loosely
packed ones since highly packed systems are less susceptible to rearrange their
conformation  when subjected to a pressure change. Thus, $\kappa_T$ 
correlates with the volume of the system~\cite{SILV98}. Now, since the volume of water is a 
convex function of temperature, $\kappa_T$ is also  convex with respect to 
temperature for water -- see Fig.~\ref{fig:water}(e). Fig.~\ref{fig:water}(f) shows that the
simulated compressibility is also a convex function of temperature, although the curvature is 
not very pronounced and its minimum is not as pronounced as in experiments.

Heat capacity, which measures the capacity of a system to store thermal energy 
($C_P=(dH/dT)_P$), is much higher in water than in ice -- see 
Fig.~\ref{fig:water}(h). This has been explained by the multiple energy storage mechanisms 
of water as the breakage of van der Waals interactions and H-bonds. The heat capacity 
of the model presents a much higher variability than real water:~close to the 
transition, $C_P$ is much higher than ice and this quantity decreases fast, reaching 
the same value as ice at about $T=0.15$.

\begin{figure}
\centerline{\includegraphics[height=60mm]{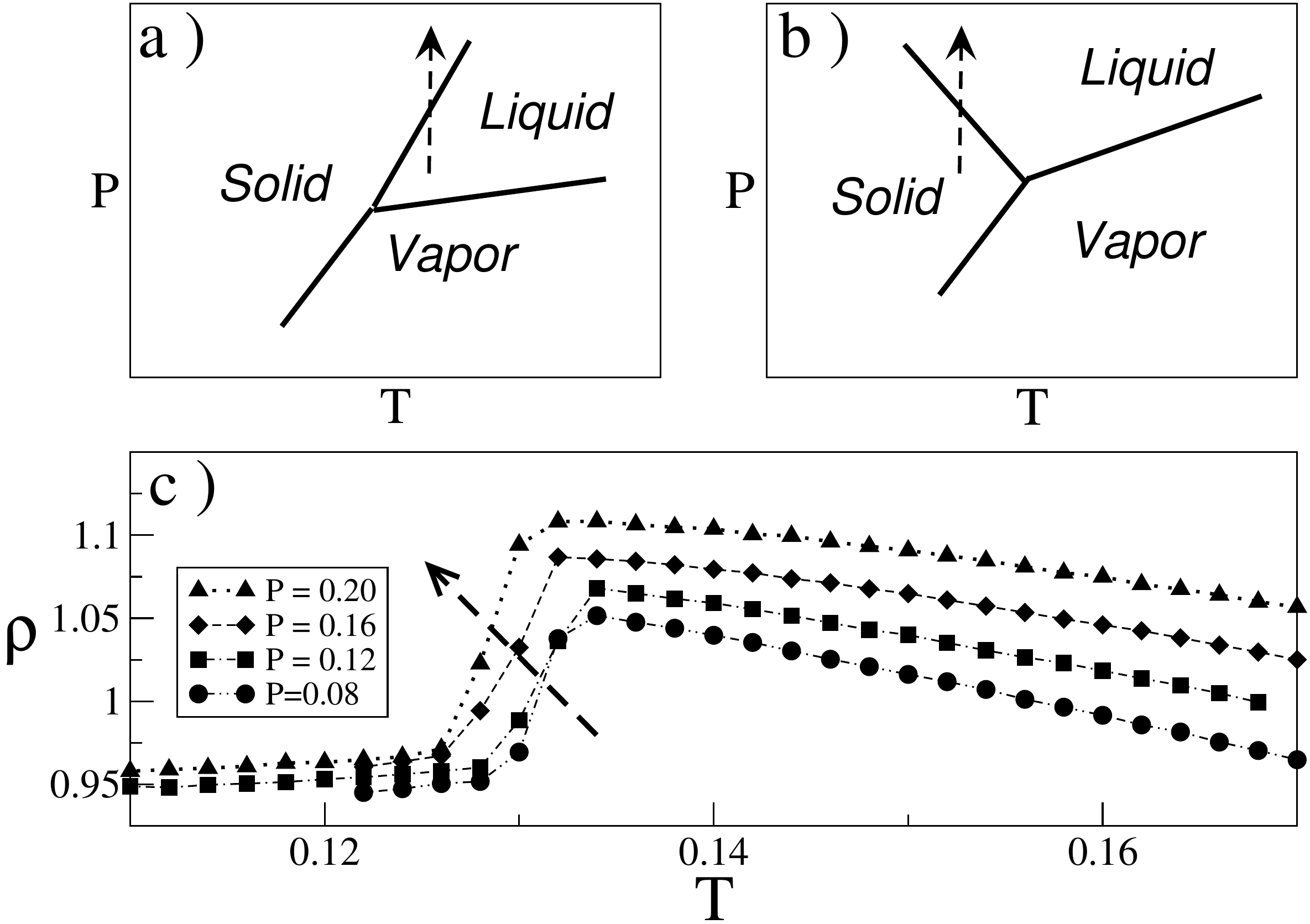} }
\caption{Schematical representation of the phase diagram of a simple one-component substance (a) and
water (b). Arrows indicate that pressure freezes a typical liquid but pressure melts 
ice. (c) Simulated density of the model for different values
of pressure. Arrow indicates the shift of the freezing temperature to lower values 
as pressure increases.}
\label{fig:clausius}
\end{figure}

In Fig.~\ref{fig:clausius}(a), we illustrate schematically the  
coexistence lines of the solid, liquid and vapor phases of a simple material. At any 
point along those lines, the free energies of the adjacent phases are equivalent and the 
Clausius-Clapeyron equation is obtained by equating them:
\begin{equation}
\Big( \frac{dP}{dT} \Big)_{coex} = \frac{\Delta h }{T \Delta v}.
\end{equation}
Since $\Delta h < 0$ and $\Delta v < 0$ for the liquid to solid transition of 
typical materials, $(dP / dT)_{coex}$ is positive. As a result of this positive slope, 
a typical liquid freezes when pressure is applied to it -- as illustrated by the arrow on
Fig.~\ref{fig:clausius}(a). On the other side, since water
expands upon freezing, $\Delta v >0$ while the enthalpy difference remains negative
(ice has a lower enthalpy compared to water). 
Thus the coexistence line of the liquid-solid transition has a negative slope, 
i.e. $(dP/dT)_{coex} < 0$. This is illustrated in Fig.~\ref{fig:clausius}(b) and leads 
to the melting of ice when pressure is applied to it. In Fig.~\ref{fig:clausius}(c) we show
that the model reproduces this anomalous behavior of water. The simulated dependence of the density
on temperature is shown for different values of pressure. The freezing temperature shifts to
lower values as pressure increases, implying $(dP/dT)_{coex} < 0$.

\begin{figure}
\vspace{-3cm}
\begin{tabular}{c}
\centerline{\includegraphics[height=145mm]{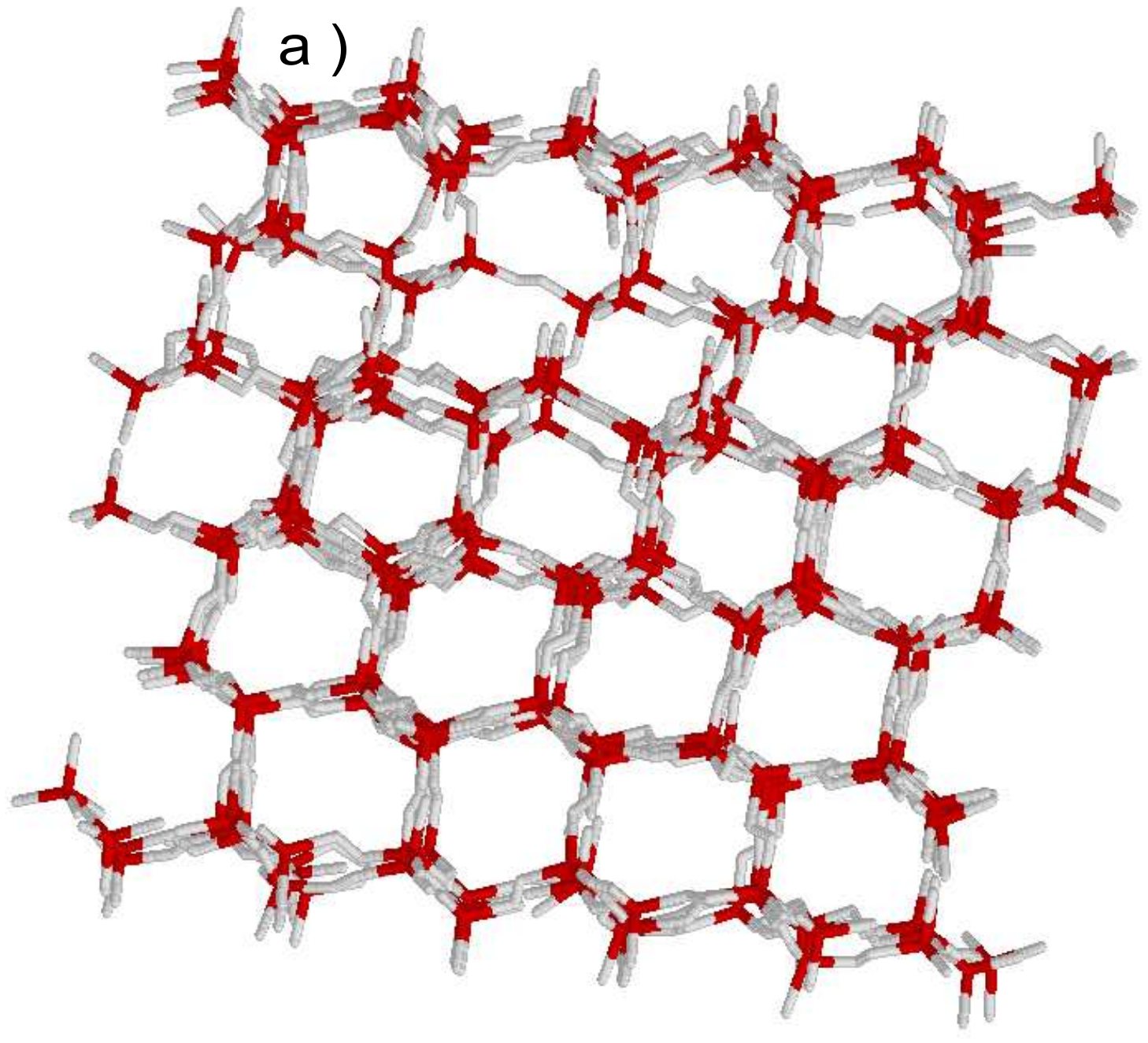} } \vspace{-1.8in} \\
\vspace{-0.5cm}\centerline{\includegraphics[height=60mm]{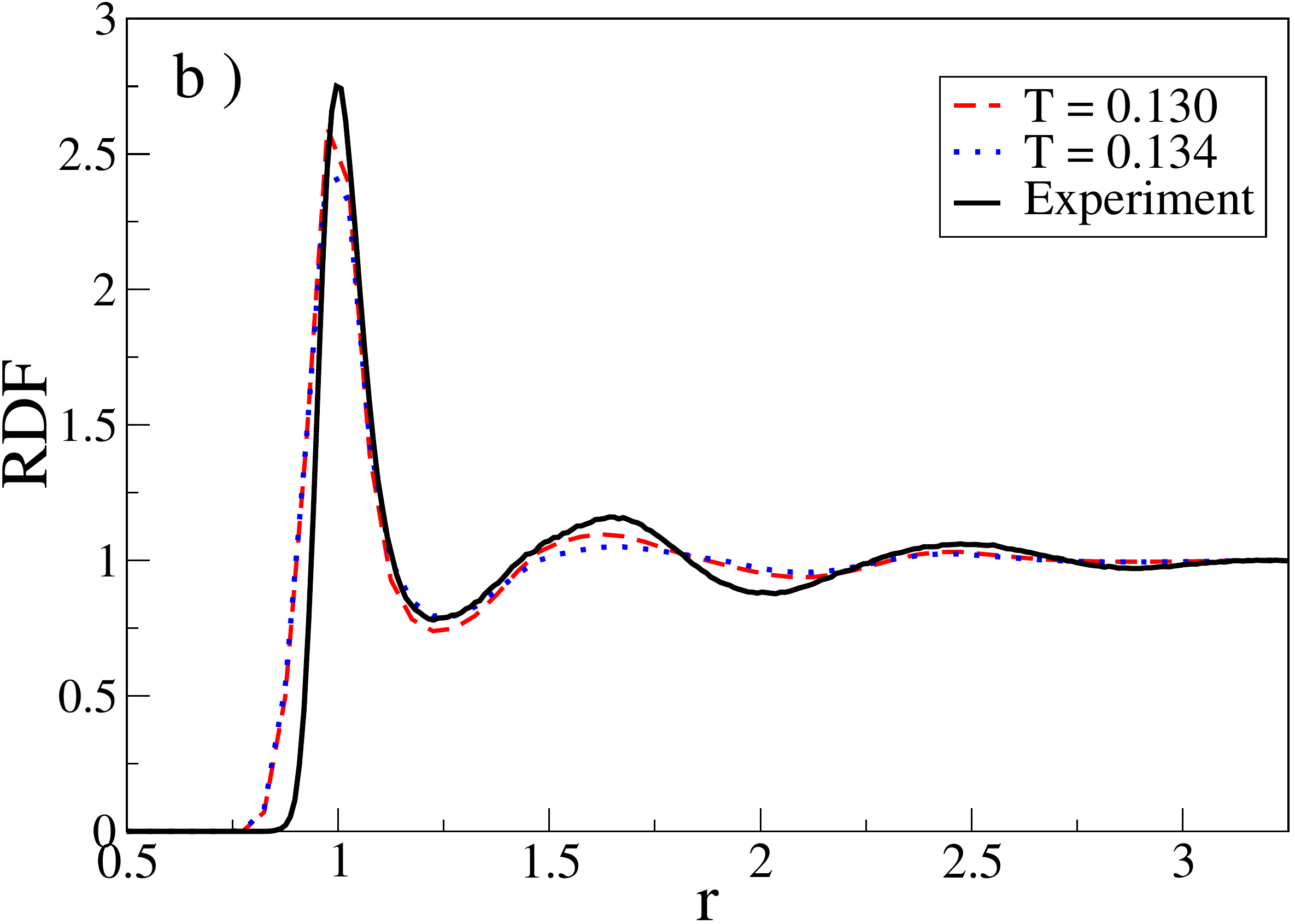} }
\end{tabular}
\caption{(Color online) (a) Ground state of the model: Ice-I where oxygen atoms occupy positions 
on a diamond-like lattice. (b) Radial distribution function of the model at different temperatures 
compared to experimental data of water at 298 K.}
\label{fig:structure}
\end{figure}

Figure~\ref{fig:structure}(a) shows the structure obtained by freezing the MB model. In this 
configuration, the center of a MB molecule occupies the sites of a diamond-like structure 
and its arms point to its four neighbors. Note that without a penalty term 
(Eq.~(\ref{eqn:penalty})) the empty spaces found in Ice-I can be the stage for the 
formation of another tetrahedral-like lattice. Thus, this term efficiently shifts the 
energy of those compact configurations and, in particular Ice-VII,
in favor of Ice-I. In Fig.~\ref{fig:structure}(b), the experimental RDF \cite{SOPE00} is compared to 
the ones of the model at different temperature. The second peak of the RDF, 
commonly referred to as tetrahedral peak~\cite{FINN01}, is a fingerprint 
of the tetrahedral geometry of water. It occurs at a distance given by the cosine 
rule, $d^2=2R^2_{HB}- 2R_{HB} \cos(109.4^{\circ})\approx 1.6$, much smaller than for a simple 
liquid~\cite{FINN01}. When compared to experiment, 
the model's RDF has slightly less structure but it peaks 
at the same position as the experiment -- indicating that the average structure of 
the model agrees well with the experiment.

\section{Conclusion \label{sec:concl}}

In this work, we have constructed a simple but realistic model for 
water based on the Mercedez-Benz approach \cite{SILV98}. At  low temperature the 
model freezes forming  Ice-I and it
reproduces the main peaks of the experimental RDF of bulk water. In
addition to these structural  properties, the model reproduces the
density anomaly of water:~ice has a lower density  than water and the
density of water is a concave function of temperature, with a maximum
close to the freezing point. Also, the slope of the solid-liquid
coexistence curve is also found to be  negative, in agreement with
experiments.

In the 2D MB model, the H-bond interaction favors environments having
3 first-neighbors at a distance $R_{HB}$ which competes
with the van der Waals interaction that favors 6 neighbors at a distance 
0.7$R_{HB}$. This competition is the underlying physics of the model 
that accounts for the density anomaly of bulk water. In the 3D MB model, 
H-bond and van der Waals interactions have the same equilibrium
distance and the density anomaly results from an energy penalty for crowded 
environments. Without this penalty, the system would solidify into a compact 
Ice-VII configuration. With the penalty term, Ice-VII conformations compete
with an open packed diamond-like structure. The competition between these 
interactions is the underlying mechanism that leads to the density 
anomaly of the system.

%

The MB model for water is based on local interactions which are much 
faster to compute than usual models that uses long-range Coulomb forces. We believe 
that this model will provide new insights into water mechanisms related to molecular
hydration. In particular, investigations of the hydrophobic effect are being undertaken with
this model.

\section*{Acknowledgements}

C.L.D. would like to thank Razvan Nistor and Marco Aurelio Alves
Barbosa for insightful discussions. He would also like to thank
Alvarro Ferraz Filho and Silvio Quezado for kindly hosting his stay at
the International Centre of Condensed Matter Physics (ICCMP) 
in Brasilia, Brazil, where 
part of this work was completed.  We would like to thank SharcNet 
(www.sharcnet.ca) for computing resources. 
M.K.  has been supported by  
NSERC of Canada and  T.A-N. 
by the Academy of Finland  through its COMP CoE and
TransPoly grants.

\bibliographystyle{unsrt}

\end{document}